\documentclass[aps,draft]{revtex4}
\usepackage[dvips]{graphicx}
\begin{document}

\title{Magnetization Cooling of an Electron Gas}

\author{ Nodar L. Tsintsadze and Levan N. Tsintsadze}
\affiliation{ Faculty of Exact and Natural Sciences, Andronikashvili Institute of Physics,  Tbilisi State University, Tbilisi, Georgia }

\date{\today}

\begin{abstract}

We propose an adiabatic magnetization process for cooling the Fermi electron gas to ultra-low temperatures as an alternative to the known adiabatic demagnetization mechanism. We show via a new adiabatic equation that at the constant density the increase of the magnetic field leads to the temperature decrease as $T\sim 1/H^2$.

\end{abstract}


\maketitle

A wide range of new phenomena arise from the magnetic field in the Fermi gas, such as the change of shape of the Fermi sphere, thermodynamics, de Haas-van Alphen \cite{haa} and Schubnikov-de Haas \cite{shu} effects, the skin effect, propagation of proper waves \cite{ltsin}, etc.
Recently, it was shown that in a strong magnetic field the Fermi gas becomes diamagnetic and the transition to the superconducting state takes place \cite{ntsin}. The study of thermodynamics of a Fermi gas of sufficiently low temperature and in a strong magnetic field is of fundamental significance for understanding physics of compact astrophysical objects, e.g., the interior of white dwarf stars, magnetospheres of neutron stars and magnetars, as well as it is of substantial interest in connection with the applications in modern technology (e.g. metallic and semiconductor nanostructures) and  next generation intense laser-solid density matter experiments. In the remote past Debye \cite{deb} and Giauque \cite{gia} independently suggested to cool materials below $1^o K$ via adiabatic demagnetization, i.e. $T\sim H$. Theoretical and experimental investigations of this effect led to the development of magnetic refrigerators.

In this Letter, we present the magnetic cooling of the quantized electron Fermi gas. This effect is based on adiabatic magnetization in contrast to the adiabatic demagnetization process for cooling of materials till ultra-low temperatures. Here we shall calculate the thermodynamic quantities in the presence of a strong external homogeneous and time independent magnetic field at the temperatures T much less or more than the degeneracy temperature $T_F$.

As well known in an electron gas at the temperature of absolute zero the electrons occupy all states with momentum from zero to a limit $p=p_F$, which $p_F=(3\pi^2)^{1/3}\hbar n^{1/3}$ is the radius of the Fermi sphere in momentum space ($\hbar$ is the Planck constant divided by $2\pi$ and n is the density of electrons).

Under the action of magnetic field H (let the magnetic field be directed along the z-axis), the electrons do not move in a straight path but rotate in circular orbits in a plane perpendicular to the magnetic field. This transverse motion is similar to the motion of linear harmonic oscillator which oscillates about an equilibrium position with the cyclotron frequency $\omega_c=\frac{e H}{m_0c}$ ($m_0$ is the electron rest mass). Such an oscillatory motion is quantized \cite{landS} and in the non-relativistic limit
the electron energy levels $\varepsilon^{\jmath,\sigma}$ are determined by the expression
\begin{eqnarray}
\label{erti}
\varepsilon^{\jmath,\sigma}=\frac{p_z^2}{2m_0}+(2\jmath+1+\sigma)\beta_B H\ ,
\end{eqnarray}
where $\jmath$ is the orbital quantum number ($\jmath=0,1,2,...$), $\ \sigma$ is the operator the z component of which describes the spin orientation $\vec{s}=\frac{1}{2}\vec{\sigma }$ ($\sigma=\pm 1$), and $\beta_B=\frac{\mid e\mid\hbar}{2m_0c}$ is the Bohr magneton.
We can see from Eq.(\ref{erti}) that in the magnetic field all allowed states within the Fermi surface are condensed on the surface of coaxial cylinders (Landau's tubes) parallel to $p_z$ axis and states lying between the orbits are forbidden.

Its also obvious from Eq.(\ref{erti}) that the energy spectrum of electrons consist of the lowest Landau level, $\jmath=0$, $\sigma=-1$, and pairs of degenerate levels with opposite polarization, $\sigma=1$. Thus each value with $\jmath\neq 0$ occurs twice, and that with $\jmath=0$ once. Therefore, $\varepsilon^{\jmath,\sigma}$ can be rewritten as
\begin{eqnarray}
\label{ori}
\varepsilon^{\jmath,\sigma}=\varepsilon^{\jmath}=\frac{p_z^2}{2m_0}+2\jmath\beta_B H=\frac{p_z^2}{2m_0}+\jmath\hbar\omega_c \ .
\end{eqnarray}

The number of quantum states \cite{landS} of a particle moving in a volume V and the interval $dp_z$ for any value of $\jmath$ is
\begin{eqnarray}
\label{sami}
\frac{2V\mid e\mid Hdp_z}{(2\pi\hbar)^2c}=\frac{V\varepsilon_F\eta m_0 dp_z}{2\pi^2\hbar^3}\ ,
\end{eqnarray}
where $\eta=\frac{\hbar\omega_c}{\varepsilon_F}$ and $\varepsilon_F$ is the electron Fermi energy.

The equilibrium total density of electrons is defined as
\begin{eqnarray}
\label{otxi}
n=\frac{m_0\varepsilon_F\eta}{2\pi^2\hbar^3}\sum_{\jmath=0}^\infty\int_{-\infty}^\infty dp_z \ f(p_z,\jmath) \ ,
\end{eqnarray}
where $f(p_z,\jmath)$ is the Fermi distribution function
\begin{eqnarray}
\label{xuti}
f(p_z,\jmath)=\frac{1}{\exp \left\{\frac{\frac{p_z^2}{2m_0}+\jmath\hbar\omega_c-\mu}{T}\right\}+1}
\end{eqnarray}
and $\mu$ is the electron chemical potential.

To evaluate the density $n$ from the expression (\ref{otxi}), we shall consider gas at the temperature limit
$\mid \jmath\hbar\omega_c-\mu\mid\gg T$. In this case the Fermi distribution function is in a good approximation described by the Heaviside step function $H(\mu-\varepsilon^\jmath)$.

Let us recall some purely quantum mechanical features of a macroscopic system. It is well known that there is an extremely
high density of levels in the energy eigenvalue spectrum of a macroscopic system. We know also that the number of levels in a
given finite range of the energy spectrum of a macroscopic system increases exponentially with the number of particles N in the
system, and separations between levels are given by numbers of the $10^{-N}$. Therefore we can conclude that in such case the
spectrum is almost continuous, and a quasi-classical approximation is applicable. Thus, we can replace the summation in Eq.(\ref{otxi}) by an integration $(\sum_1^{\jmath_{max}}\rightarrow\int_1^{\jmath_{max}}d\jmath)$ to obtain after a simple integration an expression of the density $n$. The result is
\begin{eqnarray}
\label{exvsi}
n=\frac{p_F^3}{2\pi^2\hbar^3}\left\{\eta+\frac{2}{3}(1-\eta)^{3/2}\right\}\ .
\end{eqnarray}
In equation (\ref{exvsi}) the first term is the contribution from the lowest Landau level ($\jmath=0$), i.e. this term is associated with the Pauli paramagnetism and self-energy of particles. The second one results from the summation over all higher Landau levels.

If the magnetic field is absent $(\eta=0$), then for the density we get the well known expression
\begin{eqnarray}
\label{shvidi}
n_{\mid\eta=0}=\frac{p_F^3}{3\pi^2\hbar^3}\ .
\end{eqnarray}
Whereas, if the magnetic field is very strong $\eta>1$, the sum in Eq.(\ref{otxi}) vanishes and all electrons are at the main level, which means that the gas is fully polarized and spins of all particles are aligned opposite to the magnetic field. Hence, in this case for the density we have
\begin{eqnarray}
\label{rva}
n=\frac{p_F^3}{2\pi^2\hbar^3}\ \eta\ .
\end{eqnarray}

We now use Eq.(\ref{exvsi}) to write the limiting Fermi energy and the degenerate temperature as
\begin{eqnarray}
\label{tsxra}
\varepsilon_F=T_F=\frac{p_F^2}{2m_0}=\frac{(2\pi^2)^{2/3}\hbar^2n^{2/3}}{2m_0
\left\{\eta+\frac{2}{3}(1-\eta)^{3/2}\right\}^{2/3}}\ .
\end{eqnarray}

In the case when $\eta>1$, Eq.(\ref{tsxra}) reduces to
\begin{eqnarray}
\label{ati}
T_F=\frac{(2\pi^2)^{2/3}\hbar^2}{2m_0}\Bigl(\frac{n}{\eta}\Bigr)^{2/3}\ ,
\end{eqnarray}
which reads that the degenerate temperature for the given $n$ is reverse dependent on the magnetic field.

Noting the relation $\eta=\hbar\omega_c/\varepsilon_F$ equation (\ref{ati}) can be rewritten in the form
\begin{eqnarray}
\label{tertmeti}
T_F=\gamma \Bigl(\frac{n}{H}\Bigr)^2  \ ,
\end{eqnarray}
where
\begin{eqnarray*}
\gamma=\frac{\pi^4\hbar^4}{2m_0}\frac{c^2}{e^2} \ .
\end{eqnarray*}

For the thermodynamic potential we have
\begin{eqnarray}
\label{tormeti}
\Omega=-\frac{V\mid e\mid H T}{2\pi^2\hbar^2c}\sum_{\jmath=0}^\infty\int_{-\infty}^\infty dp_z\ln
\Bigl(1+e^{\frac{\mu-\varepsilon^{\jmath}(p_z,\jmath)}{T}}\Bigr) \ ,
\end{eqnarray}
where the electron energy levels $\varepsilon^{\jmath}(p_z,\jmath)$ are determined by the expression (\ref{ori}).

Integrating over new variable $\varepsilon^{\jmath}$ and replacing the summation in Eq.(\ref{tormeti}) by an integration, from Eq.(\ref{tormeti}) we obtain
\begin{eqnarray}
\label{tsameti}
\Omega=-\frac{V(2m_0)^{3/2}\mu^{5/2}}{3\pi^2\hbar^3}\ \eta \left\{1+\frac{2}{5\eta}(1-\eta)^{5/2}\right\}-
\frac{V(2m_0)^{3/2}\mu^{1/2}T^2}{24\hbar^3}\ \eta \left\{1+\frac{2}{\eta}(1-\eta)^{1/2}\right\}\ .
\end{eqnarray}
Here we assumed that $\mid\varepsilon^{\jmath}-\mu\mid\gg T$ and $\eta < 1$.

In the case $\eta > 1$, the sum in Eq.(\ref{tormeti}) vanishes and for the thermodynamic potential we get
\begin{eqnarray}
\label{totxmeti}
\Omega=-\frac{V(2m_0)^{3/2}\mu^{5/2}\eta}{3\pi^2\hbar^3}-\frac{V(2m_0)^{3/2}\mu^{1/2}T^2\eta}{24\hbar^3}\ .
\end{eqnarray}

Having the thermodynamic potential (\ref{tsameti}) and (\ref{totxmeti}), we can define the pressure, the entropy, the specific heat, etc. for the Fermi electron gas
\begin{eqnarray}
\label{txutmeti}
P=-\frac{\Omega}{V}\ ,\hspace{1cm} S=-\Bigr(\frac{\partial\Omega}{\partial T}\Bigr)_{V,\mu} \hspace{1cm} and \hspace{1cm}
C_V=T\Bigr(\frac{\partial S}{\partial T}\Bigr)_{V,\mu} \ ,
\end{eqnarray}
which yields for the pressure
\begin{eqnarray}
\label{tekvsmeti}
P=\frac{(3\pi^2)^{2/3}\hbar^2n^{5/3}\eta}{2m_0}\ \left\{1+\frac{2}{5\eta}(1-\eta)^{5/2}\right\}+
\frac{m_0}{12\hbar^3}\ p_F T^2\eta \left\{1+\frac{2}{\eta}(1-\eta)^{1/2}\right\}
\end{eqnarray}
and for the entropy
\begin{eqnarray}
\label{chvidmeti}
S=\frac{N}{6}\ \frac{m_0}{\hbar^3}\ \frac{p_F}{n}\ T\eta \left\{1+\frac{2}{\eta}(1-\eta)^{1/2}\right\}\ ,
\end{eqnarray}
where N is the total number of electrons.

Whereas for the specific heat we have the same expression as Eq.(\ref{chvidmeti}), as it was expected.

In the absence of magnetic field, i.e. $\eta=0$, the above expressions for P, S and $C_V$ reduce to well known expressions, given in Ref.\cite{landS}.

Since the entropy remains constant in an adiabatic process, from Eq.(\ref{chvidmeti}) follows
\begin{eqnarray}
\label{tvrameti}
\frac{T\eta_0}{n^{2/3}\delta^{1/2}}\ \left\{1+\frac{2\delta^{1/2}}{\eta_0}(\delta-\eta_0)^{1/2}\right\}=const=\frac{2T_0}{n_0^{2/3}}
\end{eqnarray}
and
\begin{eqnarray}
\label{tsxrameti}
\frac{3}{2}\eta_0\delta^{1/2}+(\delta-\eta_0)^{3/2}=1\ ,
\end{eqnarray}
where $\eta_0=\hbar\omega_c/\varepsilon_{F0}$, $\delta=\varepsilon_F/\varepsilon_{F0}$, $\varepsilon_{F0}$ is the Fermi energy in the absence of the magnetic field, and $T_0$ is the initial temperature.

Note that Eq.(\ref{tvrameti}) with Eq.(\ref{tsxrameti}) is an adiabatic equation, and it should be emphasized that this adiabatic equation is the function of three variables - the density, the temperature and the magnetic field. These couple of equations were derived for the case $\hbar\omega_c<\mu=\varepsilon_F(H)$, where the expression $\varepsilon_F(H)$ is given by Eq.(\ref{tsxra}).

If we suppose that the density of the Fermi gas is constant, then the adiabatic equation reduces to
\begin{eqnarray}
\label{otsi}
\frac{\Theta}{2}\ \left\{\frac{\eta_0}{\delta^{1/2}}+2(\delta-\eta_0)^{1/2}\right\}=1
\end{eqnarray}
with the expression (\ref{tsxrameti}), where $\Theta=\frac{T}{T_0}$.

We now explicitly express the specific heat and entropy through the magnetic field for the case $\eta>1$ or $\delta<\eta_0$. With Eq.(\ref{totxmeti}) at hand, we obtain
\begin{eqnarray}
\label{otsdaerti}
C_V=S=\frac{N (2m_0)^{3/2} T\hbar\omega_c}{12\hbar^3\mu^{1/2}n}\ .
\end{eqnarray}

Taking into account Eq.(\ref{tertmeti}), which is valid for $\hbar\omega_c>\mu=\varepsilon_F$, for the adiabatic process from Eq.(\ref{otsdaerti}) we obtain a simple relation between the temperature and the magnetic field
\begin{eqnarray}
\label{otsdaori}
\frac{\Theta \emph{H}^2}{\emph{n}^2}=1\ ,
\end{eqnarray}
where $\emph{H}=\frac{H}{H_0},$ $\ H_0$ is defined by the condition
\begin{eqnarray}
\label{otsdasami}
\hbar\omega_c(H_0)=\mu=\varepsilon_F(H_0)
\end{eqnarray}
and $\emph{n}=\frac{n}{n_0}.$

For the constant density $\emph{n}=1$ of the Fermi electron gas, the relation between the temperature and the magnetic field indicates that the increase of the magnetic field consequently leads to the temperature decrease as $T\sim \frac{1}{H^2}$. Thus this is the adiabatic magnetization process for cooling the Fermi electron gas to ultra-low temperatures.

Let us estimate the threshold magnetic field $H_0$. To this end, we use Eq.(\ref{tertmeti}) and from Eq.(\ref{otsdasami}) for the metal density we find

$$ H_0\simeq 2\cdot 10^7 Gauss\ ,$$

whereas for the semiconductor ($n\sim10^{19} cm^{-3}$) we have

$$H_0\simeq 2\cdot 10^5 Gauss$$

We now consider the electron gas, which is sufficiently rarefied and at high temperature. In this case in order to calculate the entropy, we shall use the Maxwell-Boltzmann distribution function and the expression of free energy accordingly
\begin{eqnarray}
\label{otsdaotxi}
F=-NT\ln\Bigl(\frac{\ell}{N}\sum_k\ e^{-\frac{\varepsilon_k}{T}}\Bigr) \ ,
\end{eqnarray}
where $\varepsilon_k$ denotes the energy levels of a single particle and is defined by the expression (\ref{ori}).

We emphasize here that two Gibbs distribution for a variable number of particles $W_{nN}=e^{(\Omega+\mu N-E_{nN})/T}$ and $W_n=e^{\frac{F-E_n}{T}}$ $\ (F=\Omega+\mu N)\ $ are entirely equivalent except one case when the fluctuations of the total number of particles take place, and the same is true of the relation between expressions (\ref{tormeti}) and (\ref{otsdaotxi}).

Note that the expression (\ref{ori}) contains two terms, the first of which describes the kinetic energy of translational motion and the second one the discrete energy values are Landau levels.

Use of expressions of the energy (\ref{ori}) and the number of quantum states (\ref{sami}) in the sum in Eq.(\ref{otsdaotxi}) yields
\begin{eqnarray}
\label{otsdaxuti}
\sum_k\ e^{-\frac{\varepsilon_k}{T}}=\sum_{\jmath=0}^\infty\ e^{-\frac{\hbar\omega_c\jmath}{T}}\ \frac{2V\mid e\mid H}{(2\pi\hbar)^2c}\ \int_{-\infty}^\infty dp_z e^{-\frac{p_z^2}{2mT}}=2V\Bigl(\frac{m}{2\pi\hbar^2}\Bigr)^{3/2}\ T^{3/2}\ \frac{\hbar\omega_c/T}{1-
e^{-\frac{\hbar\omega_c}{T}}}\ .
\end{eqnarray}

Substituting Eq.(\ref{otsdaxuti}) into the expression of the free energy of electrons, we obtain
\begin{eqnarray}
\label{otsdaexvsi}
F=-NT\ln[2\Bigl(\frac{m}{2\pi\hbar^2}\Bigr)^{3/2}\ \ell\ \frac{T^{3/2}}{n}\ \frac{\hbar\omega_c/T}{1-
e^{-\frac{\hbar\omega_c}{T}}}]\ .
\end{eqnarray}
If the magnetic field is zero, then from Eq.(\ref{otsdaexvsi}) one gets well known expression for the free energy.

With the explicit expression (\ref{otsdaexvsi}) at hand, one can easily calculate the entropy per electron. The result is
\begin{eqnarray}
\label{otsdashvidi}
s=\frac{S}{N}=-\frac{1}{N}\frac{\partial F}{\partial T}=\ln[2\Bigl(\frac{m}{2\pi\hbar^2}\Bigr)^{3/2}\ \ell\ \frac{T^{3/2}}{n}\ \frac{\hbar\omega_c/T}{1-
e^{-\frac{\hbar\omega_c}{T}}}]+\frac{\hbar\omega_c/T}{e^{\frac{\hbar\omega_c}{T}}-1}\ .
\end{eqnarray}

We can now derive from Eq.(\ref{otsdashvidi}) the adiabatic equations. First, in the case of weak magnetic field, i.e., $\hbar\omega_c\ll T$, we have the following adiabatic equation
\begin{eqnarray}
\label{otsdarva}
\frac{T^{3/2}}{n}\left\{1-\frac{7}{12}\Bigl(\frac{\hbar\omega_c}{T}\Bigr)^2\right\}=constant.
\end{eqnarray}

Next, in the case of strong magnetic field, i.e. $\hbar\omega_c\gg T$, which is more interesting case, the adiabatic equation reads
\begin{eqnarray}
\label{otsdatsxra}
\frac{T^{1/2}H}{n}=constant\ .
\end{eqnarray}
We can see that here the adiabatic equation, similar to the case of the degenerate electron gas, contains three unknown quantities n, T and H instead of two as in the ordinary thermodynamics.

We specifically note here that the expressions of the adiabatic equations (\ref{otsdaori}) and (\ref{otsdatsxra}) in the non-relativistic limit are similar. Namely they demonstrate that the law of temperature decrease along with increase of the magnetic field at constant n is the same in both the degenerate and quasi-classical cases.

It should be also emphasized that Eq.(\ref{otsdatsxra}) requires the temperature of the electron gas to satisfy two conditions $H>10^4\cdot T$ and $T>T_F$, where $T_F$ is the Fermi energy and it is determined by the equation (11) of Ref.\cite{ltsin}.

To summarize, we have investigated the effects of the quantization of the orbital motion of electrons and the spin of electrons on the thermodynamic quantities in an electron gas in the non-relativistic limit. Namely, we have derived the pressure, the specific heat and the entropy taking into account the spin and the quantization of the orbital motion of the electrons. We specifically note here that all thermodynamic quantities obtained by us reduce to well known expressions when the external magnetic field goes to zero. We have obtained a novel adiabatic equation, which implies that at the constant density the increase of the magnetic field consequently leads to the temperature decrease as $T\sim 1/H^2$ . Thus, we have demonstrated the adiabatic magnetization process for cooling the Fermi electron gas to ultra-low temperatures. In addition, we have shown the similar law of temperature decrease along with increase of the magnetic field at constant n in the Boltzmann, i.e. the quasi-classical limit. The results of the present paper may attract interest from scientists and companies to develop new kinds of magnetic refrigerator designs. This refrigeration, once proven viable with increasing energy efficiency, could be used in any possible application where cooling, heating or power generation is used, including in spacecraft.


\begin{references}

\bibitem{haa} W.J. de Haas and P.M. van Alphen, Proc. Netherlands Roy. Acad. Sci. {\bf 33}, 1106 (1930).

\bibitem{shu} L.W. Schubnikov and W.J. de Haas, Proc. Netherlands Roy. Acad. Sci. {\bf 33}, 130 (1930).

\bibitem{ltsin} L.N.Tsintsadze, AIP Conf. Proc. {\bf 1306}, 89 (2010); e-print arXiv: physics/0911.0133v1; e-print arXiv: physics/1005.3408v1.

\bibitem{ntsin} N.L.Tsintsadze and L.N.Tsintsadze, to be published.

\bibitem{deb} P. Debye, Ann. d. Phys. {\bf 81}, 1154 (1926).

\bibitem{gia} W.F. Giauque, Journ. Amer. Chem. Soc. {\bf 49}, 1864 (1939).

\bibitem{landS} L.D.Landau and E.M.Lifshitz, {\sl Statistical Physics}, Part 1 (Pergamon Press, Oxford, 1989) 175.


\end{references}
\end{document}